\documentclass[twocolumn,english,aps,prl,superscriptaddress]{revtex4-1}
\usepackage[latin9]{inputenc}
\usepackage{amsmath}
\usepackage{amssymb}
\usepackage{graphicx}
\usepackage{esint}
\usepackage{color}
\usepackage[usenames,dvipsnames]{xcolor}

\makeatletter
\@ifundefined{textcolor}{}
{%
 \definecolor{BLACK}{gray}{0}
 \definecolor{WHITE}{gray}{1}
 \definecolor{RED}{rgb}{1,0,0}
 \definecolor{GREEN}{rgb}{0,1,0}
 \definecolor{BLUE}{rgb}{0,0,1}
 \definecolor{CYAN}{cmyk}{1,0,0,0}
 \definecolor{MAGENTA}{cmyk}{0,1,0,0}
 \definecolor{YELLOW}{cmyk}{0,0,1,0}
}

\newcommand{\rs}{\rm\scriptscriptstyle}

\makeatother

\usepackage{babel}
\begin{document}

\title{Disordered BKT transition and superinsulation}

\author{S. Sankar}

\affiliation{Department of Theoretical Physics, Tata Institute of Fundamental
Research, Homi Bhabha Road, Navy Nagar, Mumbai 400005, India}

\author{V. Vinokur}

\affiliation{Materials Science Division, Argonne National Laboratory, 9700 S.
Cass. Ave, Lemont, IL 60437, USA.}

\author{V. Tripathi}

\affiliation{Department of Theoretical Physics, Tata Institute of Fundamental
Research, Homi Bhabha Road, Navy Nagar, Mumbai 400005, India}

\begin{abstract}
Strongly disordered superconducting films have been observed to undergo finite temperature transitions to a superinsulating state, of apparently infinite resistance,
mirroring superconductivity. Approaching the transition, some of the films reportedly exhibit Berezinskii-Kosterlitz-Thouless (BKT) criticality implying that superinsulation 
is associated with an ordered charge BKT phase. An even more singular Vogel-Fulcher-Tammann (VFT) criticality has also been seen, positing the question of the existence of 
fundamentally different states of finite temperature insulators. Here we develop a theory of the criticality of a disordered lateral Josephson junction array with weak 
Josephson coupling. We show that it is equivalent to a two-dimensional Coulomb gas subject to a random potential with logarithmic correlations.
We show that strong disorder results in a 
regime exhibiting VFT criticality instead of the usual BKT one, and find that it corresponds to transition to a nonergodic insulator phase.
\end{abstract}
\maketitle

More than 40 years ago, celebrated works by Berezinskii, Kosterlitz, and Thouless (BKT) introduced the idea of topological phase transitions 
where pairs of bound topological vortex-like excitations unbind at the critical temperature\,\cite{Berezinskii1970,Berezinskii1971,Kosterlitz1972,Kosterlitz1973}. 
Once it was recognized that BKT theory applies to two-dimensional (2D) superconductors and planar Josephson junction arrays (JJA), the realization came that in 2D systems 
true superconductivity is a manifestly low-temperature BKT phase, where vortices are bound in vortex dipoles, see A.\,M.\,Goldman in\,\cite{Jose2013}  
and also\,\cite{Baturina2012}. Above the BKT transition temperature $T_{\rs BKT}$,
the proliferation of low-energy vortices breaks down the global phase coherence, and the 2D superconducting system enters a resistive state. 
Studying the JJA, Fazio and Sch\"{o}n demonstrated that depending on the ratio of the Josephson coupling, $E_{\rs J}$ to the Coulomb energy of a single junction, $E_{\rs C}$, the JJA would host either superconductivity, 
if $E_{\rs J}/E_{\rs C}\gg 1$, or an insulator in the opposite limit, where physics is dominated by the charge degrees of freedom\,\cite{fazio1991}. They showed that if
 the junction capacitance $C\gg C_0$, where $C_0$ is the capacitance 
to the ground, the electrostatic screening is suppressed and charges interact logarithmically over large distances. Then the vortex-charge duality transformation 
$v_i\leftrightarrow q_i$, $\pi E_{\rs J}\leftrightarrow2E_{\rs C}/\pi$ retains the form of the action describing the JJA, with $q_i$ being the charge of the junction 
$i$ and $v_i$ being the vorticity of the corresponding plaquette in JJA. The zero-temperature superconductor-to-insulator transition (SIT) occurs at the self-dual 
point $E_{\rs J}/E_{\rs C}=2/\pi^2$ and the charge binding-unbinding BKT transition at the insulating side takes place at $T_{\rs BKT}=E_{\rs C}/2$ (throughout the text we use the energy units for temperature taking Boltzmann constant $k_{\rs B}=1$).   

The next important step came with Diamantini and co-workers\,\cite{perugio} who constructed a gauge theory of JJA at zero temperature having revealed striking 
consequences of the charge -- vortex symmetry. They established that due to logarithmic confinement of charges 
on the insulating side of the SIT, vortices condense into a 
superfluid condensate and the novel state with the infinite resistance (i.e. a mirror dual image of a superconducting state with the infinite conductance) that they 
termed a \textit{superinsulator} forms.  Two years later, Doniach\,\cite{Doniach1998} having utilized the duality between 
vortices and Cooper pairs\,\cite{fisher1990}, independently introduced superinsulating phase of 2D superconductors as a phase where vortices form 
a superfluid condensate. The latter blocks the charge motion thus giving rise to vanishing of the electric conductivity.

The final understanding of the nature of the superconductor-to-insulator transition came in 2008, when the experimental discovery of current jumps in strongly 
disordered superconducting films\,\cite{shahar2004,Baturina2007} lead to the realization that disordered superconducting films in the critical region of the SIT 
are endowed with a diverging dielectric constant and thus can harbor 2D logarithmic 
Coulomb interactions between the Cooper pairs over appreciably macroscopic scales\,\cite{Fistul2008}. With recognition that logarithmic 
interaction between charges is a physically realizable realm, the earlier concept of the vortex-Cooper pairs duality\,\cite{fisher1991} 
came to full fruition and lead to the concept of a superinsulator as a confined charge BKT state manifesting the symmetry of the 
uncertainties in the phase and charge of the Cooper condensate which compete according to the Heisenberg principle\,\cite{vinokur2008superinsulator,BVAnnals2013}. 

The superinsulator proposal in\,\cite{vinokur2008superinsulator} ignited an explosive interest and has been attracting ever since the intense attention of researchers,
see\,\cite{BVAnnals2013,shahar2015} and references therein. 
A great deal of effort was expended to identify the charge BKT transition as a precursor of the superinsulation. There appeared a tantalizing report of the critical Vogel-Fulcher-Tammann 
(VFT) like behavior, $R_{\square}(T)\sim \exp[{\mathrm{const}/(T-T_{\rs{VFT}})}]$, of the sheet resistance of the InO film upon freezing into the finite temperature insulator\,\cite{shahar2015}, resembling (but more singular than) the BKT 
dependence $R_{\square}(T)\sim \exp[{\mathrm{const}/\sqrt{T-T_{\rs{BKT}}}}]$\,\cite{Kosterlitz1973}.
The latter was found in experiments on NbTiN films\,\cite{Baturina2016}. 
These findings raise important questions: (i) What is the physical mechanism of the VFT behavior and (ii) what, if any, is the connection between the 
BKT and VFT-like divergences of the resistance? In Ref. \cite{shahar2015}, the finite temperature insulator was attributed to the 
many body localization (MBL) mechanism\,\cite{Gornyi2005,Basko2006}.
In this Letter we unravel the origin of the VFT and establish its close relation to the BKT physics. 

\begin{figure}
\centering
\includegraphics[width=.95\columnwidth]{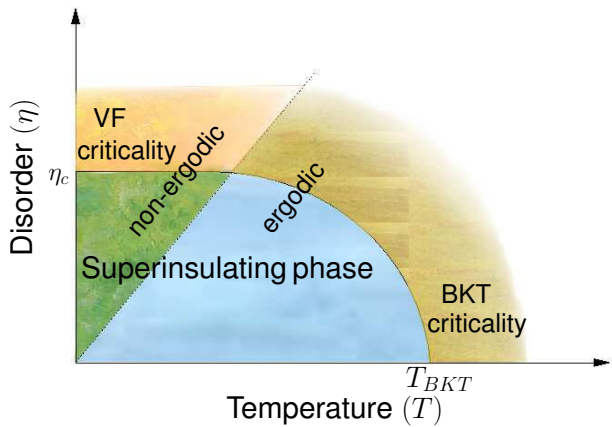}
\caption{\label{fig:phasediag} A sketch of the phase diagram of the superinsulating state and critical behaviours 
of a two-dimensional Josephson-junction disordered array in disorder-temperaure coordinates. 
Disorder being considered is the quenched random dipole moments of the grains. In the
superinsulating phase, the probability of single charge excitations is zero. The transition to the
conducting state occurs via the proliferation of the single charge excitations generated either
thermally or by disorder. The former leads to the BKT criticality, given by Eq.\,(1), while the
latter results in VFT behavior of Eq.\,(2). The dotted line $\eta = T/2E_{\rs C}$, separates the nonergodic
region (shaded green), where the charge dipoles freeze (their free energy becomes independent
of temperature), from the ergodic region (shaded blue) where a finite entropy is associated with
the charge dipoles which can appear anywhere. Likewise, the VFT critical region is nonergodic
and conducting, while the BKT critical region is ergodic and conducting.}
\end{figure}

We adopt a planar JJA model taking into account the effect of 
quenched random dipole moments of the superconducting grains and show that for weak Josephson coupling, $E_{\rs J}\ll E_{\rs C}$, it is equivalent to a Coulomb gas 
subject to a disordered potential with logarithmic long-range correlations. We demonstrate that for sufficiently strong disorder, the usual
BKT critical behavior turns into a more singular VFT criticality characteristic to glasses\,\cite{Anderson1979,Palmer1984}. In this glassy phase, the charge 
excitations freeze into a nonergodic insulator phase. Weakly disordered systems retain their BKT criticality and freeze into an ergodic insulator phase. 
Namely, if the system state lies below the ergodic-nonergodic transition line $T^*(\eta)$ in $T$-$\eta$ coordinates (see Fig.\,\ref{fig:phasediag}), 
where $\eta$ measures 
the strength of the charge dipole disorder, then the conductivity follows the standard BKT criticality, 
 \begin{equation}
 \sigma(T)\sim e^{-\text{const}/\sqrt{T-T_{\text{ BKT}}}}\,.
 \label{BKT}
 \end{equation}
 and above the transition line, the system exhibits the VFT critical behavior,
 \begin{equation}
 \sigma(T)\sim e^{-\text{const}/(T-T_{\text{VFT}})}\,.
 \label{VF}
 \end{equation}
%
 %a definite signature of the  nonergodic superconducting phase.
 %
 Based on these findings, we propose that the 
recent observations of the VFT critical behavior of conductivity in disordered InO superconducting thin films\,\cite{shahar2015} and the
charge BKT criticality proposed for
  %TiN\,\cite{BVAnnals2013}
  %and 
  NbTiN\,\cite{Baturina2016} films on approaching the finite temperature insulator transition are 
  respective manifestations of nonergodic and ergodic insulator phases.
  We indeed find some resemblance between the discovered ergodic vs. nonergodic BKT behaviors and
those in the MBL picture\,\cite{altshuler2016} also showing both BKT\,\cite{Gornyi2005} and VFT\,\cite{Sarang2014} criticality, but, as we will see further, there are 
critical differences between them. 

%The 
%concept of the BKT transition went beyond the Landau paradigm, establishing a continuous phase transition mediated by topological excitations
%but occuring without spontaneous symmetry breaking. 
 %The BKT ideas have had a profound impact on the physics of
 %a broad spectrum of systems ranging from superconductors to ultra-cold atoms and the quark confinement problems\cite{Kosterlitz2016,Jose2013,Greensite2011}. 
 
We consider a Cooper pair insulator (CPI) that forms on the insulating side of the SIT in superconducting films. 
%Possible sources of this disorder include deviations
%from stoichiometry, interstitials and vacancies. 
Coarse-graining over the size of the Cooper pairs, we approximate the disorder background charge distribution 
$\rho(\mathbf{r})$ as a Gaussian white noise correlation function, 
$\langle (\rho(\mathbf{r}) - \bar{\rho})(\rho(\mathbf{r}') - \bar{\rho}) \rangle = n_d\delta(\mathbf{r} - \mathbf{r}'),$ 
where the average background charge density, $e\bar{\rho}$, equals the average charge density of the CP, and $n_d$ is the variance of the coarse-grained 
background charge distribution. The angular brackets stand for averaging over disorder. 
Near the SIT the dielectric constant diverges $\kappa \gg 1$, see Ref.\,[\onlinecite{BVAnnals2013}] and references therein, 
and in a film of thickness $t$, 
the Coulomb interaction between two charges has logarithmic separation $r$ dependence as $\ln(\Lambda/r)$ over distances $t < r <\Lambda$, where 
$\Lambda\simeq\kappa t$ is an electrostatic screening length\,\cite{Rytova1967,BVAnnals2013}. 
At distances beyond $\Lambda$ the interaction falls off as~$1/r$.
In disordered films, a CPI is customarily viewed as a lateral JJA comprising superconducting droplets coupled by Josephson links\,\cite{BVAnnals2013}. %see Fig.\,\ref{fig:granular}. 
The  
droplets nucleate at deep potential fluctuations resulting from intrinsic quenched charge disorder of the host.
%Hence JJA in an insulating state appears a generic model for CPI.
Near the SIT, the size of the droplets is expected to be of order of the superconducting coherence length and in any case to exceed %determined by the competition
%of disorder and interparticle interactions greatly exceeds 
the characteristic localization length of single particles in the 
disordered potential\cite{larkin1995,falcoprb,sarath2016}.
In the JJA, the effective dielectric
constant and, accordingly, the crossover length is expressed via the characteristic capacitances\,\cite{fazio1991}. 
We will address the situation $\Lambda\gtrsim L$ so that the interactions between the charges is logarithmic. 

The excess charge on a droplet interacts with the charge distribution of other droplets. 
The leading contribution to the energy is provided by the electric `monopoles', the single excess charges $n_{\mathbf i}$ on the other droplets, 
$-\sum_{\mathbf{i \neq j}} E_{\rs{C}}n_{\mathbf i}n_{\mathbf j}\ln|({\mathbf r}_{\mathbf i}-{\mathbf r}_{\mathbf j})/a|,$ where $a$ is a microscopic length scale 
(the size of the droplet), $E_{\rs C} = q^2/2C$ is the characteristic energy for creation of a CP dipole ($q=2e$) across neighboring droplets and $C$ is the inter-droplet 
capacitance\,\cite{fazio1991}. The next order contribution comes from the dipole moments of 
the grains, $\mathbf{P_{i}}$, which yield the random potential energy
\begin{equation}
  V_{\mathbf{i}}=\sum_{j}\frac{q}{2\pi C}\frac{\mathbf{P}_{j}\cdot\mathbf{r}_{ij}}{r_{ij}^{2}}.
\label{Vrand}  
\end{equation}
Using $\langle\mathbf{P}\rangle=0$,
we derive, analogously to\,\cite{petkovic2009}, that the dipole-induced random potential is logarithmically correlated:
\begin{equation}
\langle(V(\mathbf{r})-V(\mathbf{r}^{\prime}))^{2}\rangle \approx 4 \eta E_{\rs C}^{2} \ln \left({|\mathbf{r}-\mathbf{r}^{\prime}|}/{R}\right),
\end{equation}
where $\eta=\pi\langle P^{2}\rangle/q^{2}R^{2}$, $R$ is the typical radius of a grain and $\langle P^2 \rangle \propto n_d.$

%%%%%%%%%%%%%%%%%%%%%%%%%%%%%%%%%%%%%%%%%%%%%%%%%%%%%%%%%%%%%%%%%%%%%%%%%%%
%%%%%%%%%%%%%%%%%%%%%%%%%%%%%%%%%%%%%%%%%%%%%%%%%%%%%%%%%%%%%%%%%%%%%%%%%%%
The effective action for JJA comprises both, the charge and phase degrees of freedom. Trading off the phase degrees 
of freedom for vortex variables via the Villain transformation\,\cite{fazio1991}, we obtain the Fazio-Sch\"{o}n action 

\begin{align}
S[n,v] & =\int_{0}^{\beta}d\tau\sum_{\mathbf{i},\mathbf{j}} \biggl( E_{\rs C}n_{\mathbf{i}}U_{\mathbf{ij}}n_{\mathbf{j}}
+E_{\rs J}v_{\mathbf{i}}U_{\mathbf{ij}}v_{\mathbf{j}}+\iota n_{\mathbf{i}}\Theta_{\mathbf{ij}}\partial_{\tau}v_{\mathbf{j}} \nonumber \\  
  & \qquad\qquad +\frac{1}{2E_{\rs J}}\partial_{\tau}n_{\mathbf{i}}U_{\mathbf{ij}}\partial_{\tau}n_{\mathbf{j}} \biggr)+\sum_{i}V_{\mathbf{i}}n_{\mathbf{i}},\label{charge_vortex_action}
\end{align}
where $v_{\mathbf{i}}$ are the integer-valued vortex degrees of freedom,
defined on the dual lattice, $\Theta_{\mathbf{ij}}=\arctan\left(\frac{y_{i}-y_{j}}{x_{i}-x_{j}}\right)$ and $U_{ij}=-\ln|\mathbf{r}_{i}-\mathbf{r}_{j}|$.
In the insulating state where 
$E_{\rs J}\ll E_{\rs C}$, $E_{\rs J}$ being the typical strength of the Josephson coupling, we can treat the integer valued vortex
fields as continous fields and integrate out them to obtain the effective charge action as
\begin{equation}
S_{e}[n]=\int_{0}^{\beta}d\tau\sum_{\mathbf{i,j}}U_{\mathbf{ij}}\left(\frac{1}{E_{\rs J}}\partial_{\tau}n_{\mathbf{i}}\partial_{\tau}n_{\mathbf{j}}+E_{\rs {C}}n_{\mathbf{i}}n_{\mathbf{j}}\right)+\sum_{i}V_{\mathbf{i}}n_{\mathbf{i}}\,.\label{eq:dynamical-charge_model}
\end{equation}
Hereafter we neglect the temporal fluctuations as they are irrelevant at low energies and do not alter the nature of the phase transition governed by the interplay
of the long-range Coulomb interaction and disorder correlations. 
The critical behavior of the resulting classical 2D Coulomb gas Hamiltonian subject to a random potential with logarithmic correlations
is obtained by analyzing the disorder averaged real-space Kosterlitz renormalization group (RG) equations.
Following Ref.\,\cite{carpentiernucphysb} we introduce replicas and then perform the average over
disorder to obtain the averaged replicated Coloumb gas Hamiltonian (with $m$ replicas),
\begin{align}
\beta H^{(m)}=\sum_{i\neq j}K_{ab}n_{i}^{a}\ln\left(\frac{|\mathbf{r}_{i}-\mathbf{r}_{j}|}{a_{0}}\right)n_{j}^{b}+\sum_{i}\ln Y[\mathbf{n}_{i}].
\end{align}
Here the superscripts on the charges refer to the replica index, $Y[\mathbf{n}]=\exp(-n^{a}\gamma K_{ab}n^{b})$ is the fugacity,
$K_{ab}=\beta E_{\rs {C}}\delta_{ab}-\eta\beta^{2}E_{\rs {C}}^{2}$ is the effective coupling, and $a_{0}$ is of the order of the
lattice constant and serves as a short length cutoff as we go over to the continuum description.
Significant contribution to the partition function only comes from
charges $\pm1,0$ and hence we resctrict to these.
To $O(Y[\mathbf{n}]^{2})$, one obtains the following RG flow equations for the effective coupling and fugacity as we rescale from 
$a_{0}(\ell)$ to $a_{0}(\ell+d\ell)= a_0e^{d\ell}$ (thus, $a_0 e^{\ell}$
is the spatial scale over which short wave-length excitations have been integrated out
by RG):
\begin{align}
\partial_{\ell}(K_{\ell}^{-1})_{ab} & =2\pi^{2}\sum_{\mathbf{n}\neq 0}n^{a}n^{b}Y[\mathbf{n}]Y[-\mathbf{n}]\label{eq:renorm_coupling}\\
\partial_{\ell}Y[\mathbf{n}\neq 0] & =(2-n^{a}K_{ab}n^{b})Y[\mathbf{n}]+\sum_{{\bf n}^{'}\neq 0,{\bf n}}\pi Y[\mathbf{n}^{'}]Y[\mathbf{{\bf n}-n}^{'}].\label{eq:renorm_fugacity}
\end{align}
Equation (\ref{eq:renorm_coupling}) comes from the annihilation of
dipoles of opposite vector charges in the annulus $a_{0} <|r_{i}-r_{j}|<a_{0} e^{d\ell}$.
Simple rescaling gives the first part of Eq.\,(\ref{eq:renorm_fugacity}).
The second part comes from the possibility of ``fusion'' of unit charges in two different replicas 
upon coarse graining. After taking the appropriate analytic continuation in the $m \rightarrow 0$ limit \cite{carpentiernucphysb} (also see SI), 
one finds that the RG equations can be recast in terms of the distribution function, $P_{\ell}(z)$, of the scale dependent single charge fugacity $z$. 
Introducing the generating functional for $P_{\ell}(z)$ as 
\begin{equation}
G_{\ell}(x)=1-\int_{-\infty}^{\infty}du\tilde{P}_{\ell}(u)\exp[-e^{\beta(u-x+E_{\ell})}],\nonumber
\end{equation}
where  $u=1/\beta\ln(z)$, the distribution $\tilde{P}(u)$ is defined as, $\tilde{P}(u)\text{d}u=P(z)\text{d}z$, 
and $E_{\ell}=\int_{0}^{\ell}E_{\rs C}(\ell^{'})d\ell^\prime$, one reduces the RG equation
to the compact form of the  Kolmogorov-Petrovsky-Piskounov (KPP) equation for the generating functional
 \begin{equation}
 \partial_\ell G_{\ell} = (\eta E_{\rs C}^2)\partial_{x}^{2}G_{\ell} + 2 G_{\ell}(1-G_{\ell})\,,
 \label{eq:KPP}
\end{equation}
(see SI).

The phase boundary between the `superinsulating' and `normal insulating' phases can be inferred from the behavior of $P_{\ell}(1)$ defined by the KPP equation. 
We identify two distinct regimes:
 (i) $P_{\ell}(1)$ decreasing with $\ell$ corresponding to the superinsulating phase with $\sigma(T)=0$, and (ii) $P_{\ell}(1)$ increasing with $\ell$ describing the 
 `normal insulating phase' with activation, finite albeit exponentially small,  $\sigma(T)\neq 0$. 
 The phase boundary corresponds to stationary $P_{\ell}(1)$ and is given by
 \begin{align}
  2-\frac{E_{\rs C}}{T}+\frac{\eta E_{\rs C}^{2}}{T^{2}}  = 0 & \mbox{ for } T>T_{g}=E_{\rs C}\sqrt{\frac{\eta}{2}}; \label{eq:11}\\
  \eta  =\eta_{c}=\frac{1}{8} & \mbox{ for } T \le T_{g}, \label{eq:12}
  \end{align} %\\ \mbox{ and } \\
where $E_{\rs C}$ and $\eta$ stand for the renormalized quantities at $l=\infty$. 
We immediately identify two distinct critical behaviors on approach to the charge BKT transition. 
Near the phase boundary at small degrees of disorder, the correlation length exhibits the BKT criticality

\begin{equation}
 \xi \sim e^{1/\sqrt{b/[(T/T_{\rs BKT})-1]}},\,\, (T-T_{\rs BKT})/E_{\rs C}\ll 1\,,
 \label{eq:xi-KT}
\end{equation}
where, $T_{\rs{BKT}} = E_{\rs C}/2$\,\cite{fazio1991} is the critical temperature of the charge-BKT transition 
and $b$ is a numerical constant of order unity. For finite but small disorder $\eta$, the dependence of $T_{\rs{BKT}}$ on $\eta$ can
be obtained from the solution of Eq. (\ref{eq:11}).
Near the disorder-controlled phase boundary, the correlation length is (see SI):
\begin{equation}
\xi \sim e^{1/(\eta(T)-\eta_{c})},\,\, T/E_{\rs C}\ll 1,\,\eta-\eta_{c}\ll1\,,
\label{eq:xi-disorder}
\end{equation}
where $\eta_{c}=1/8$ is the critical disorder strength at low temperatures for the transition.

These distinct critical scenarios correspond to system freezing into either the ergodic, at low disorder, or into the nonergodic, at strong disorder, respectively, 
superinsulating states as shown in the phase diagram in disorder-temperature coordinates in the Fig.\,\ref{fig:phasediag}. 
The dotted line %in Fig.~\ref{fig:phasediag1}, 
$\eta^{\ast}(T) =T/2E_{\rs C}$, marks the onset of freezing of charge dipoles\,\cite{tang1996}
where freezing means that the free energy of dipole excitations loses an explicit temperature dependence. At this line the entropy and 
hence the free energy of a dipole, has a singularity,
indicating that $\eta=\eta^{\ast}(T)$ is the phase boundary between the two distinct phases of the superinsulator.
\iffalse To reveal their nature, we adopt the procedure devised in\,\cite{tang1996} and map the process of the formation of the 
low-lying energy states in the superinsulator onto the energy of a directed polymer
passing through a sequence of nodes of a Cayley tree\cite{note}. As long as disorder is weak, there exists a 
multitude of trajectories spanning nearly the whole space that gives nearly equal energy contribution, hence the ergodic character of the phase where 
$\eta<\eta^{\ast}$. In contrast, at $\eta>\eta^{\ast}$, there is a unique ``path'' of the directed polymer along which the electrostatic energy is minimized. This is a
highly nonergodic situation since this path does not explore a finite fraction of the space. Hence the region $\eta>\eta^{\ast}(T)$
is the nonergodic state of a superinsulator. The distinct screening behaviors in the ergodic and nonergodic phases is illustrated in Fig.\,\ref{fig:screening} of 
charges by the dipoles in the corresponding cases.\fi In the ergodic phase the dipoles can appear anywhere and thus assume the most efficient -- for screening -- 
configuration. In  the nonergodic phase, the dipoles are frozen, as they emerge mostly due to fluctuations in the random quenched potential, and hence may not provide an 
efficient screening as compared to the one due to thermally generated dipoles. This then leads to the more singular VFT-like critical behavior.

The experimentally measurable quantity is
conductivity, $\sigma\simeq\mu_{c}n_{c}$, where $\mu_{c}$ is the charge mobility 
and $n_{c}\sim1/\xi^{2}$ is the density of free charges in the critical regime\,\cite{halperin1979}. Then Eq.\,(\ref{eq:xi-KT}) leads us to our result in Eq.\,(\ref{BKT}).
Turning to the strong disorder case described by Eq.\,(\ref{eq:xi-disorder}), 
we assume an activated temperature dependence 
$n_{d}(T) = n_d(0) +  N_{d}e^{-E_{d}/T}$, where $E_{d}$ is
the characteristic dopant-carrier binding energy for dopant levels near the conduction or valence bands. 
Let $T_{c}$ be the temperature at which $\eta(T_{c})=\eta_{c}$.
Expanding $n_{d}(T)$ in the vicinity of $T_{c},$ $n_{d}(T)\approx n_{d}(T_{c})[1+(T-T_{c})(E_{d}/T_{c}^{2})]$,
we recover the VFT law for conductivity near the \textit{disorder}-driven transition with $T_{{\rs VFT}}=T_c$
and $\mathrm{constant}=2T_{c}^{2}/(E_{d}\eta(T_{c}))$.
Note that this result is obtained under the condition $T_{c}<E_{\rs C}$,
for otherwise the condition for the thermally-driven BKT transition
is satisfied first with the increasing temperature and one obtains the BKT behavior of Eq.\,(\ref{BKT}) 
as expected for the weak disorder case.
Comparing VFT and BKT results one concludes that
the transition from the VFT to the BKT behavior occurs at $\eta(T_{\text{ BKT}})=\eta_\mathrm{c}$. Since critical behaviors of strongly and weakly disordered CPI belong 
in different universality classes, it is natural that they freeze into two distinct phases of the superinsulator. 

A far reaching implication of our findings is that the observation of the critical behavior may serve as an experimental indicator of whether the CPI freezes into the either 
ergodic or nonergodic superinsulating state. Based on the existing data, one can suggest that disordered superconducting 
TiN and NbTiN films\,\cite{vinokur2008superinsulator,BVAnnals2013,Baturina2016} exhibit ergodic freezing into the superinsulator, while the VFT criticality 
reported in InO films\,\cite{shahar2015} may manifest nonergodic behavior.

Our analysis applies equally to the superconducting transition where disorder that breaks time reversal symmetry is induced, for example, by random 
Aharanov-Bohm phases on the links\,\cite{petkovic2009}, or by paramagnetic impurities. 
A key signature of the nonergodic superconducting phase would be the resistivity critically vanishing according to the 
VFT law. Instead of tuning the temperature, one can also tune an external (perpendicular) magnetic field to control the disorder parameter $\eta$ (see Ref.\,\cite{sarath2016}). 
The same critical scaling now appears in the magnetic field dependence of resistance near the field tuned superconductor-insulator transition.

Note that the obtained critical behaviors preceding the formation of the superinsulating state are to some extent paralleled by the critical behavior of $\sigma(T)$ arising in the MBL framework\,\cite{Gornyi2005,Basko2006,Sarang2014}.
The notable difference is that while the MBL phase is essentially nonergodic, the superinsulating phase is ergodic for $T>2\eta E_{\rs C},$ and non-ergodic for 
$T< 2\eta E_{\rs C}$. In the nonergodic region, the transition from the superinsulating phase to the conducting phase is reminiscent of the transition to 
nonergodic conducting regime derived in the MBL framework\,\cite{altshuler2016}. It is noteworthy that MBL studies have been focusing so far on the case with the short range 
interactions, while the extension of the MBL results including long-range interactions is disputable, see Ref.\,[\citenum{fleishman1980}]. 
At variance, our analysis crucially relies on the long range logarithmic nature of the interactions and the accompanying logarithmically correlated disorder.
The comparison of BKT and MBL pictures is summarized in Table 1 in SI.

To conclude, our findings establish the phase diagram of the
intriguing superinsulating state and shed new light on the superconductor-insulator transition critical region. 
More work now is required to unravel the details of the interplay between the glassiness, ergodicity and quantum coherence which play key roles in the critical region 
of strongly disordered supercondctors.  

\textit{Acknowledgments--} The work was supported by the U.S. Department of Energy, Office of Science, Materials Sciences and Engineering Division (V.M.V.)
 and by Department of Science and Technology, Govt. of India, through a Swarnajayanti grant (no. DST/SJF/PSA-0212012-13) (V.T.).

\end{document}